\documentclass[aps,prl,twocolumn,groupedaddress,showpacs,floatfix]{revtex4}
\usepackage{amsmath}
\usepackage{amssymb}

\newcommand{\dd}{\text{d}}

\newcommand{\ee}{\text{e}}

\newcommand{\p}{\partial}

\newcommand{\br}{\text{\bf r}}

\newcommand{\eps}{\varepsilon}

\usepackage{graphicx}
\begin{document}
\title{Universality class of nonequilibrium phase
transitions with infinitely many-absorbing-states}
\author{Fr\'ed\'eric van Wijland}
\email[]{Frederic.van-Wijland@ th.u-psud.fr}
\affiliation{P\^ole Mati\`ere et Syst\`emes Complexes (CNRS FR2438, Universit\'e
de Paris VII) and Laboratoire de physique th\'eorique (CNRS UMR8627), Universit\'e de Paris-Sud,
91405 Orsay cedex, France}

preprint LPT Orsay 02-47

\date{\today}

\begin{abstract}
We consider systems whose steady-states exhibit a nonequilibrium phase
transition from an active state to one --among an infinite number-- absorbing state, as some
control parameter is varied across a threshold value. The pair contact process,
stochastic fixed-energy sandpiles, activated random walks and many other
cellular automata or reaction-diffusion processes
are covered by our analysis. We argue that the upper critical dimension below
which anomalous fluctuation driven scaling appears is $d_c=6$, in contrast to a
widespread belief (see \cite{dickman-resume} for an overview). We provide the exponents governing the critical
behavior close to or at the transition point to first order in an $\eps=6-d$
expansion.
\end{abstract}

\maketitle
\noindent {\it Absorbing-state phase transitions. }
Nonequilibrium phase transitions between an active and an absorbing state are
encountered in a variety of fields ranging from chemical kinetics to the
spreading of computer viruses~\cite{reviews}. From a theoretical standpoint
absorbing state transitions form natural
counterparts to equilibrium phase transitions. The transition rates used in the
stochastic dynamics employed to model the
physical phenomenon under consideration do not satisfy detailed balance (with
respect to an {\it a priori} defined  energy function). In spite of
this apparent freedom, the number of universality classes that the transition
can fall into is incredibly small. Among known universality classes, that of
directed percolation (DP) is by far the broadest. And indeed, in the absence of
additional symmetries or conservation laws, as was conjectured twenty years ago
by Grassberger~\cite{grassberger79}, an absorbing state transition will invariably fall into the DP
universality class. Only if the dynamics possess additional features such as
particle number parity conservation, or the coupling to an auxiliary field
(whether conserved or not, static  or diffusing, {\it etc.}), or the existence
of an infinite number of absorbing states, will the transition belong to a new class. Apart from the formal interest
in classifying nonequilibrium phase transitions and in identifying which
microscopic ingredients make a transition belong to a given universality class,
there exists a greater challenge. As recently summarized by
Hinrichsen~\cite{hinrichsen}, in spite
of its domination in the theoretical physics literature, the directed percolation
univerality class was actually never observed in a real experiment. This is
often
attributed to the presence of defects in a real experiment (DP is known to be very
sensitive to quenched disorder) or to other ill-controlled effect, such as a
hidden conservation law or a coupling to an auxiliary field.

On one hand the interest in absorbing state transitions was recently enhanced by the
discovery by Vespignani, Zapperi and coworkers~\cite{fes} of their
relationship with the
ubiquitous phenomenon of self-organized criticality (SOC)~\cite{jensen98}. It was established that
the scaling behavior observed there was entirely governed by an underlying
phase transition (thereby, incidentally, tempering the mystics of SOC). On the
other hand, in a
separate wave of articles, research has focused on absorbing-state transitions
in which the order parameter freezes into one among an infinity of absorbing
states, but without any additional conservation law. The paradigmatic example of
a system showing such a behavior is the pair contact process, initially
introduced by Jensen and Dickman~\cite{dimer}, for which Mu\~noz and
coworkers~\cite{munozetal} devised a convincing phenomenological picture that we shall later use
as our starting point. 

The existence of an infinite number of absorbing states (in the large-system
limit) and the coupling to an auxiliary  static field are
the common characteristics to the microscopic models that we wish
to investigate here. It is the purpose of the present work to
provide a full renormalization group picture of the phase transition at work in
systems possessing an infinite number of absorbing states, with or without an
additional conservation law. We shall
rely on a phenomenological Langevin approach as a starting point. From there
we shall show how renormalization group arguments can be applied. This will lead
us to finding the upper critical dimension of those models. Then we will sketch
the reasoning leading to the expression of the critical exponents within the
framework of an expansion around the upper critical dimension. As will be clear
there is interesting new physics
to learn from the
many technical obstacles that pave the way to the full scaling picture.\\

We now turn to a presentation of two microscopic models chosen for their
representativeness and ease of formulation in which the absorbing state phase transitions we wish to study
appear. We also introduce the Langevin equations encoding their dynamics.
Then we sketch the field-theoretic line of reasoning leading to the
computation of the critical exponents. Finally we provide a critical
discussion of our results in the light of the existing literature.\\

\noindent {\it Two models and their phenomenological coarse-grained
description.} 
In the Pair Contact Process (PCP), particles are thrown on a lattice. A pair 
may either annihilate
$A+A\to\emptyset$ or produce a single offspring $A+A\to A+A+A$. Each lattice
site can be occupied by at most one particle. Since isolated particles cannot
diffuse, a configuration in which pairs of nearest neighbors are absent remains
frozen in time. The order parameter $\psi$ is local density of pairs of nearest neighbor
particles and the control parameter is the branching over annihilation rate
ratio. The Langevin equation believed to describe the dynamics of $\psi$ was
coined by Mu\~noz {\it et al}~\cite{munozetal}. It reads
\begin{equation}\begin{split}\label{PCP-Langevin}
\text{PCP:}\;\;\p_t\psi(\br,t)=&D\Delta_{\br}
\psi(\br,t)-\sigma\psi(\br,t)-g_1\psi^2(\br,t)\\&-g_3\psi(\br,t)\ee^{-\int_0^t\dd
t'\psi(\br,t')}+\eta(\br,t)
\end{split}\end{equation}
with $\eta$ a Gaussian white noise whose correlations are $\langle\eta(\br,t)\eta(\br',t')\rangle=g_2
\psi\delta^{(d)}(\br-\br')\delta(t-t')$. The coefficients $\sigma, g_1,g_2$ and
$g_3$ are
coarse-grained analogs of the reaction rates. The memory term is the
signature of the feedback of isolated particles on the pair dynamics. For a
detailed explanation of the physical origin of the various contributions
appearing in Eq.~(\ref{PCP-Langevin}) we refer the reader to \cite{munozetal}. The list of microscopic models whose
coarse-grained description is believed to be encoded in the Langevin
equation (\ref{PCP-Langevin}) also includes the Transfer Threshold
Process~\cite{mendes94} and various models for
catalysis (the dimer reaction~\cite{dimer}, the dimer-dimer~\cite{dimer-dimer} or the
dimer-trimer models~\cite{dimer-trimer}). A detailed analysis of the mean-field properties of this equation was provided
by Mu\~noz {\it et al.}~\cite{munozetal}.\\ 

The second family of models that we are interested in is embodied by the
so-called Manna or stochastic fixed-energy sandpiles (FES)~\cite{manna,fes,bigfes}. Grains of energy are
initially randomly distributed on a
lattice. Whenever a lattice site is occupied by more than $2$ particles (in
dimension $d$) the excess particles randomly hop to a nearest neighbor site. The
number of active sites plays the role of the order parameter. The total number of
particles is strictly conserved. An evolution equation for the local density of
active sites $\psi$ was recently proposed by Dickman, Vespignani {\it et al.}~\cite{fes}. It reads
\begin{equation}\begin{split}\label{FES-Langevin}
\text{FES:}\;\;\p_t\psi&=D\Delta_{\br}\psi(\br,t)-\sigma\psi(\br,t)-g_1\psi^2(\br,t)\\
&+g_4\psi(\br,t)\int_0^t\dd
t'\;\Delta_{\br}
\psi(\br,t')+\eta(\br,t)
\end{split}\end{equation}
with $\eta$'s correlations having the same expression as in the PCP case and the
coefficients $\sigma,g_1,g_2,g_4$ are coarse-grained analogs of the microscopic
transition rates which depend on the conserved quantity. The
nonlocal memory term expresses that space fluctuations of the static and conserved
field have a feed-back effect upon the order parameter dynamics. Again a rich variety of models were shown to be described by the same
coarse-grained dynamics, such as the Conserved Transfer Threshold
Process,
Activated Random Walkers~\cite{rpv00}, or some model for epidemic spreading in which healthy
individuals are static~\cite{epidemic}.\\

There are some hazards in relying on na\"{\i}vely built phenomenological
equations as many other interactions are generated by a coarse-graining
procedure. In principle, all terms allowed by symmetries should be included. We
shall later see that, indeed, some relevant symmetry-preserving terms have to be
considered. Let us now recall the common mean-field behavior of those
models. Denoting for both models by $\sigma$ the deviation of the control
parameter from its critical threshold value, in the steady-state the order parameter behaves as
\begin{equation}
\psi
\left\{
\begin{array}{ll}\propto|\sigma|^\beta,&\sigma\to 0^-\\
=0,&\sigma\geq 0\end{array}
\right.
\end{equation}
In mean-field we simply have $\psi=-\sigma/g_1$. At the critical point, the order
parameter decays according to
\begin{equation}
\psi(t)\sim t^{-\delta},\;\;\;t\text{ large}
\end{equation}
The correlation length and the relaxation time diverge as $\sigma\to 0$
according to $\xi\sim |\sigma|^{-\nu}$ and $\tau\sim |\sigma|^{-\nu z}$,
respectively. Within the framework of a mean-field analysis one finds the
following values for the critical exponents: $\beta=1$, $\delta=1$, $\nu=\frac
12$ and $z=2$.\\ 

\noindent {\it Analytic strategy.} The interplay between short-time and short-distance fluctuations with long
range correlations lies at the heart of the anomalous scaling observed in the
vicinity of a critical point (``anomalous'' to be understood as non mean-field).
Short distance singularities (usually referred to as ultraviolet divergencies) govern the way scaling properties deviate from their
mean-field expressions. The usual strategy is to retain the leading UV
divergencies and to perfrom a renormalization group analysis with those only. In the vast majority
of cases, this is sufficient to reach physical conclusions. Sometimes however,
contributions that 
were thrown away in the course of the analysis are crucial in preserving the
correct physics, though they are irrelevant in determining the renormalization
group fixed point. This is the present situation.\\

Let us start with the PCP. The picture is the following. We expand the
exponential memory term and find that the contribution $g_3\psi\int_0^t\psi$ exhibits the leading short-time and short-distance
singularity. Power counting shows that the bare dimension of $g_2 g_3$
is $d-6$ (in units of a length). This type of interaction is known to describe the dynamical
percolation universality class~\cite{JanssenDyP, cardygrassberger}, which has an
upper critical dimension $d_c=6$. Yet, in the present case, the phase diagram 
would not be reproduced correctly if only the truncated expansion of the exponential
were kept. As seen on the mean-field evolution equation, and as
already noted many times in the literature~\cite{munozetal}, the asymptotics are governed by the
local nonlinear term $-g_1\psi^2$ in Eq.~(\ref{PCP-Langevin}). However $g_1 g_2$
has bare dimension $d-4$ which signals that it is in fact a dangerously
irrelevant coupling, along with the subsequent powers of the argument of the
exponential. The coupling $g_1$ is not relevant in determining the
underlying fixed point structure but essential in preserving the overall phase
diagram. An important consequence is that the usual scaling assumption for the
order parameter 
\begin{equation}\label{echelle}
\langle\psi(t)\rangle=b^{-\frac{d+\eta}{2}}{\cal
F}(b^{-z}t,b^{1/\nu}|\sigma|),\;\;b\text{ large}
\end{equation}
must be abandoned since the scaling function ${\cal F}$ will exhibit a singular behavior in
the $g_1$ variable as the latter approaches 0. In mean-field ${\cal F}$ depends
on $g_1$ as $1/g_1$. While irrelevant variables are
traditionnally omitted from the list of arguments of ${\cal F}$, in the present
case this would lead to unphysical conclusions. The next step to see how this dangerously irrelevant coupling $g_1$ is
renormalized to 0 in the vicinity of the dynamical percolation fixed point (a
necessity overlooked in~\cite{munozetal}). In
order to do this we have followed the technical procedure recalled by Janssen
and Schmittmann~\cite{JanssenSchmittmann} (see also the references
therein). We skip all technical steps~\cite{raf}, and merely quote the final
result: $g_1(b)\sim b^{-2-\frac{\eps}{7}}$ as the coarse-graining
scale $b$ is increased. We have denoted by $\eps=6-d$ and the result for the
exponent is given to first order in $\eps$. In order to obtain the leading
$\eps$ behavior we have combined the mean-field expression for the scaling function
${\cal F}$ appearing in Eq.~(\ref{echelle}) as far as its  $g_1(b)$ dependence is
concerned with the scaling properties of the field, time and $\sigma$ at the dynamical percolation
fixed-point. This has led us to the
following critical exponents,
\begin{equation}\label{expcritiques}
\beta=1-\frac{3}{14}\eps,\;\;\delta=1-\frac{1}{4}\eps,
\end{equation}
the expressions of which are given to leading order in $\eps=6-d$.

As far as the FES described by Eq.~(\ref{FES-Langevin}) are concerned, the situation is a bit more delicate. A rather
involved
RG analysis~\cite{raf} shows that in fact the dynamical percolation vertices are generated
already at one-loop order (the one-loop graph obtained by connecting $g_2$ and
two $g_4$'s leads to effective $g_1$ and $g_3$ vertices, and no effective $g_4$). This is because, as is very often  the case with
gradient interactions, the short-scale behavior of the memory term
$-g_4\psi(\br,t)\int^t_0\dd t'\Delta_{\br} \psi(\br,t')$ converts into an
effective $-g_3\psi(\br,t)\int_0^t\dd
t'\psi(\br,t')$ contribution (and higher powers) after coarse-graining. This
also explains the failure of naive power counting directly on
(\ref{FES-Langevin}) which, naively performed, would lead to the erroneous conclusion that $d_c=4$
since $g_2 g_4$ has bare dimension $d-4$ while $g_2 g_3$ has dimension $d-6$. And then we may apply the reasoning of the previous
paragraph and the results of Eq.~(\ref{expcritiques}) continue to hold for
FES.\\

What the field-theory approach teaches us can be summarized as follows. First, a
detailed analysis of the renormalized  interactions shows that both
classes of models considered are described by the same-field theory which has an upper-critical
dimension $d_c=6$. This result is in contradiction with the existing literature
of the last ten years~\cite{dickman-resume}. Critical exponents are expected
to behave differently from mean-field in space dimensions $d<6$. We have
determined the critical exponents of the phase transition to leading order in
$\eps=6-d$. As to FES-like systems, at the renormalization flow fixed point, the conservation law is
irrelevant. This seemingly innocuous property leads to technical difficulties since the
conservation law is crucial in saving the overall phase diagram. In both the PCP
and the FES cases,  while the underlying fixed point is that of dynamical percolation,
computing the critical exponents requires to perform an independent renormalization of an
irrelevant coupling. Thus the critical exponents cannot be deduced from those of
percolation by means of hyperscaling relations. The validity of
our approach holds in the vicinity of $d_c=6$. As $d$ is lowered, already at $d=4$ new qualitative
features show up in the theory. Since the percolation fixed point becomes
trivial in $d=1$ an entirely new picture must inevitably set in as dimension
is lowered, perhaps as early as $d$ is decreased below $d=4$ and possibly
again at $d=2$. Splitting of the universality classes of the PCP and FES in low
space dimensions must be
envisaged as well.\\

There are several criticisms that can be opposed to the present work. First of all,
the use of phenomenological Langevin equations sometimes proves hazardous, but,
fortunately other formalisms~\cite{ft} allow exact mappings onto field-theories having
exactly the same features as the ones discussed in this work. Indeed, the
effective Langevin equation (\ref{FES-Langevin}) used for sandpiles is, strictly
speaking, not correct, since coarse-graining (loop corrections) shows that other dominant terms are
generated. Most
importantly, however, we must admit that there is absolutely no numerical
evidence supporting our findings. An obviously too concise summary of the numerical
state of affairs is that both FES~\cite{julien} and PCP belong to
the DP class in dimension one (recent studies disagree with the results of
\cite{julien}, such as \cite{dickmanend=1}). But as dimension is increased PCP is still found
to belong to the DP class while FES are found to form an independent
universality class. L\"ubeck, in some recent simulations~\cite{lubeck}, claims
that $d_c=4$ by performing simulations directly in high dimensions. A cheap way out for the theoretician is to refer the reader
to a recent preprint by Grassberger~\cite{grassbergerpreprint} in the somewhat different but related context of
forest-fires in
which dynamical percolation plays some role as well. There 
numerical proof is provided  that it might well be
impossible with present day computers to ever reach the true asymptotic scaling
regime. We recall that there are many obstacles on the numerical side: the
impossibility to use simple finite-size scaling relations (due to the presence
of a dangerously irrelevant operator). Quenched
disorder (for FES~\cite{rq}) and other long-term memories are known to be difficult to overcome
numerically, and finally the proximity of the directed percolation
fixed point whose influence must be felt until the system eventually crosses
over to its actual asymptotics. All of those features, we fear, play a part in rendering
the reaching of the true asymptotics a hopeless endeavor. 
As far as the observed lack of universality of spreading exponents is
concerned~\cite{munozetal}
we believe that the phenomenon can be understood within a renormalization group
picture. As demonstrated in a much simpler case in \cite{wij98} spreading exponents
exhibit nonuniversal values that depend on the moments of the initial
distribution of both the order parameter and auxiliary fields. This dependence
is all the stronger as the initial distribution deviates from
a strict Poissonian law.
In less favorable cases the initial distribution introduces couplings at the
initial time that cannot even be renormalized, thus questioning the existence of a
scaling regime.\\

%%%%%%%%%%%%%%%%%%%%%%%%%%%
As a conclusion we summarize our findings. We have provided field-theoretic
arguments showing that a large number of
stochastic processes exhibiting a phase transition between an active state and
an absorbing state (picked up among an infinite number of such) belong to the universality class of dynamical percolation and
therefore have $d_c=6$ as their upper critical dimension (instead of $d_c=4$ as
appears in the last decade literature). The existence of a
static field to which the order parameter is coupled is the common feature to
all considered models. We have provided
expressions for the critical exponents within the framework of  renormalized
perturbation theory in the vicinity of the upper critical dimension to leading
order in $\eps=6-d$. A technical account of the results presented in
this letter is in preparation~\cite{raf}.\\

\begin{acknowledgments}The author akcknowledges several discussions with HK Janssen, HJ Hilhorst, A Vespignani, R.
Pastor-Satorras, M.A. Mu\~noz, H. Chat\'e and J. Kockelkoren.
\end{acknowledgments}


\begin{thebibliography}{10}
\bibitem{rq}
The initial static field fluctuations play the same role as quenched
disorder added to a simple DP process~\cite{JanssenDPdisorder}, which cannot
easily be accounted for in a Langevin equation for $\psi$ only.
\end{thebibliography}
\end{document}